# From relativistic to quantum universe: Observation of a spatially-discontinuous particle dynamics beyond relativity

**Sergey A. Emelyanov**

Ioffe Institute, 194021 St. Petersburg, Russia

Correspondence: sergey.emelyanov@mail.ioffe.ru

**Abstract**. We perform an experimental test where we directly observe light-induced electron transitions with a *macroscopic* spatial discontinuity. The effect is related to the fundamental indivisibility of macroscopic orbit-like quantum states reminiscent of so-called extended states in the integer quantum Hall system. The test has become realizable due to the discovering of a quantum phase with spontaneous pervasive quantum ordering reminiscent of that of a single atom. The observed transitions may be regarded as a peculiar quantum dynamics beyond relativity, which implies that the current relativistic model of universe should be replaced by a deeper quantum model. It is the Bohm's model of undivided universe which now should involve a deeper-than-classical concept of absolute simultaneity and a deeper-than-relativistic concept of space and time. Ultimately, our test thus establishes a new hierarchy of fundamental physical theories where the de Broglie-Bohm realistic quantum theory is the deepest theory which does not contradict either classical physics or relativity but rather is beyond both. This is because the fact that quantum theory is dealing with a deeper reality where physical objects are not self-sufficient entities and therefore their discontinuous transitions are possible within an overall quantum system which may well be macroscopic.

## I. Introduction

### 1.1. The current view of the universe: why the Minkowski model of spacetime

As it follows from the history of science, our view of universe is not unchangeable but determined by the physical theory which is regarded as being the most fundamental in a given historical period. For example, our view of space and time is determined by the laws of spatial dynamics, which follow from this theory. Before the beginning of the twentieth century, the most fundamental theory was the classical physics. In the frames of this theory, the universe was regarded as a collection of physical bodies that evolve in space and time so that they may influence each other only through the so-called interactions. Their spatial dynamics was regarded as being always continuous and therefore it always can be characterized by the notion "speed". The concept of absolute simultaneity was based on the lack of any fundamental restrictions on the absolute value of speed and accordingly space and time were regarded as being two independent categories.

However, at the beginning of the twentieth century, our view of universe changed drastically as the other physical theory began to be regarded as the most fundamental. It is the theory of relativity, which, on the one hand, remains the classical view of universe as a collection of interacting physical bodies but, on the other hand, establishes a strict limitation on the absolute value of speed, which fundamentally cannot be higher than the speed of light. As a result, as the notion of absolute simultaneity is based on the notion "speed", we are compelled to come to the notion of a relative simultaneity and hence to a new insight of the structure of universe. It is the well-known Minkowski model where time is rather an additional coordinate very similar to the spatial coordinates. And the characteristic feature of this model is the fundamental impossibility of any nonlocal signaling as it inevitably leads to the violation of causality.

## *1.2. Bohr's quantum theory: why "there is no quantum world…"*

Today, when the Minkowski model remains the basis of our insight of universe for more than a hundred years, the "taboo" on a nonlocal signaling is generally regarded as a God-given thing as it is generally believed that no spatial dynamics may occur beyond relativity. But surprisingly this belief appears at stake almost immediately after the recognition of the Minkowski model because almost in parallel with relativity a one more fundamental physical theory emerges which introduces a completely new physical concept known neither in classical physics nor in relativity. It is the quantum theory with the concept of a fundamental discontinuity of physical processes including the spatial dynamics. Indeed, if we regard realistically such quantum effect as the electron transitions between the atomic orbitals, then this process looks like a completely new spatial dynamics beyond relativity. Moreover, as such, quantum formalism does not restrict the lengthscale of spatial discontinuity. Rather, all restrictions are related to the lengthscale of the system where the transitions occur. Therefore, if we suppose that there are no principle differences between the microcosm and the macrocosm, then these restrictions do not look fundamental.

As known, the problem with quantum transitions was solved in a very extraordinary way, that is, through the recognition of Bohr's interpretation of quantum theory, which ultimately makes illegal any rational view of quantum effects in accordance with the Bohr's principle "*There is no quantum world. There is only a quantum mechanical description of physical phenomena.*" In fact, this interpretation implies that there are two different worlds: a non-existent (or mystical) quantum world and the realistic world of our everyday experience. With this approach, quantum particles are not the objectively-existing physical bodies and we can meaningfully talk about them only in the context of a specific measurement carried out by an observer in the real world. Accordingly, quantum formalism may not have a physical meaning as it is only a mathematical algorithm that magically allows us to predict the outcome of an experiment carried out in the real world.

*1.3. Paradoxes of "non-existent" quantum world*

Although Bohr's approach seems inevitable, it gives rise to a number of logical paradoxes. This fact was noted even by the founding fathers of quantum theory and, first of all, by Albert Einstein who formulates one of such paradoxes through the following phrase: *"I like to think the moon is there even if I am not looking at it."* Actually, he thus draws attention to the simple and reliable fact that any macroscopic object (for example, the moon) ultimately consists of a great number of quantum particles. Therefore, if these particles exist only insofar as they are a subject of observation, then the moon should also exist only insofar as somebody is looking at it. In fact, this paradox is unresolved even up to now. At best, physicists are ready to recognize the problem with the so-called quantum-to-classical transition. But neither Bohr himself nor his followers can explain in a consistent way where is the infinitely-sharp boundary which separates the non-existent quantum world from our real world as their overlapping seems absurd.

A one more opponent of Bohr's theory was Erwin Schrödinger who pointed out that the concept of two physical worlds contradicts the fact that actually these worlds are inseparable in the sense that we potentially can design a macro-system which, for some reasons, would obey the quantum laws and hence its behavior should be interpreted in terms of the non-existent quantum world. And this may lead to sheer absurdity. As an example, Schrödinger proposes a *gedanken* experiment known today as the paradox of Schrödinger's cat. In this experiment, both a cat and a closed bottle with poison are in the same chamber. In this case, if we design a mechanism such that the state of the bottle (broken or not) is determined by a purely quantum process, say, by a radioactive decay, then the state of the cat (alive or dead) should also be interpreted in terms of Bohr's theory. This means that we should recognize that in the absence of an observer the cat is alive and dead simultaneously and moreover in a well-defined proportion. And if somebody opens the chamber and finds the cat dead, then he (or she) may be regarded as the murderer of this innocent creature because a more fortunate observer could find a living cat. in a good health.

The also opponent of Bohr's theory is John Bell. From his extensive criticism, we take only one argument expressed through the following rhetorical question: *"Was the world wave function waiting to jump for thousands of millions of years until a single-celled living creature appeared? Or did it have to wait a little longer for some more highly qualified measurer – with a Ph.D.?"* [1]. In fact, Bell thus shows that the very notion "observer" is not properly defined in

Bohr's theory though this notion plays a key role in his postulates. Moreover, it seems that this notion cannot be defined even in principle at least because there are a number of unobservable cosmic objects and it seems extremely unlikely that no quantum processes occur there.

In principle, the list of paradoxes can be continued. But all of them cannot refute the fact that quantum theory is known as the most successful theory in the history of science. This means that we do not know any cases where quantum theory would give an incorrect prediction and moreover the accuracy of its predictions is really amazing. For this reason, the rejection of this theory is absolutely inconceivable and the only question is that whether or not a realistic version of quantum theory is possible?

### *1.4. Latent progress of quantum realism: From de Broglie's seminal ideas to the Bohm's model of undivided universe*

Historically, a realistic version of quantum theory was discussed even in the early days of this theory, that is, at the Solvay Conference of 1927 [2]. We mean the so-called pilot-wave theory presented by Louis de Broglie. At the Conference, however, this theory was strongly criticized by some participants and therefore the author eventually abandons further efforts in this direction. A second birth of this theory took place much later (in 1952) and is connected with the name of David Bohm [3-4]. In fact, Bohm presented a complete realistic version of quantum theory, the predictions of which are fully consistent with the predictions of Bohr's theory at least in the experimentally-accessible situations. As one would expect, Bohm's work was highly appreciated by most adherents of physical realism. De Broglie, for example, expressed an intention to come back to his own idea. As for John Bell, he was even more enthusiastic and wrote literally the following: *"But in 1952 I saw the impossible done… Bohm showed explicitly how parameters could indeed be introduced, into nonrelativistic wave mechanics, with the help of which the indeterministic description could be transformed into a deterministic one. More importantly… the subjectivity of the orthodox version, the necessary reference to the "observer", could be eliminated…"* [5].

But one of the main achievements of Bohm's works is that they actually develop a new "quantum" view of universe, which is dictated by quantum laws if they are perceived realistically. In fact, all his subsequent articles and monographs are a complete presentation of this view which differs in principle from the view that underlies both classical physics and relativity [6-8]. The basic difference is that we should give up the centuries-old mechanistic view of universe as a collection of independent physical bodies which may interact with each other. This view is actually inconsistent with the quantum concept of fundamental indivisibility. Instead, we should come to a deeper understanding which implies that our universe is actually an

*indivisible whole*. Bohm (together with Basil Hiley) expresses this idea as follows: "*…We have reversed the usual classical notion that the independent "elementary parts" of the world are the fundamental reality, and that the various systems are merely particular contingent forms and arrangements of these parts. Rather, we say that inseparable quantum interconnectedness of the whole universe is the fundamental reality and that relatively independent behaving parts are merely particular and contingent forms within this whole...*" [6]. Thus, the "message" of quantum theory appears to be such that there is a pervasive interconnection between physical bodies, which is much deeper than what is known as "interactions" and which encourages them to behave as an indivisible whole. Strictly speaking, this indivisible whole is ultimately the only fundamental thing in the physical world. Accordingly, what we call wavefunction (or pilot wave, or quantum potential etc.) is actually a characterization of the indivisibility of physical objects. It therefore differs drastically from any classical waves at least because it does not have a source, does not carry an energy, may be undamped at large distances and cannot be detected by any devices. To characterize a spatial ordering related to this wave, Bohm introduces the notion of an *implicate order* to distinguish it from an explicit spatial ordering characterized by the Cartesian grid which actually is suitable only when we are dealing with the classical situation of quite separable physical objects.

It is easy to see that the Bohm's "quantum" view of physical world allows one to interpret realistically those facts which seem a-priori mystical in the frames of the "classical" view which inevitably leads to the Bohr's idea of "non-existent" quantum world. Indeed, as we see, both Bohm and Bohr regard quantum particle as a non-fundamental entity. Actually, this fact directly follows from the so-called contextuality of the properties of quantum particles as well as from their ability to exercise spatially-discontinuous transitions. But the clear advantage of Bohm's view is that it allows one to interpret quite realistically the non-fundamentality of quantum particles. For example, the contextuality of their properties may be interpreted in terms of their fundamental inseparability from the measuring instruments and therefore what we call their properties is actually the properties of an overall quantum system. Accordingly, spatially-discontinuous electron transitions may be interpreted in terms of the behavior of an overall quantum system such as the atom. Thus, the clue to a realistic interpretation of quantum phenomena lies in the fact that we should give up the mechanistic view of quantum particles as fundamental physical bodies, i.e. something like microscopic "building blocks" of the universe. Instead, we should start with a fundamental indivisibility of the universe and then analyze whether or not some of its constituent parts may be regarded independently in a given situation.

Thus, through Bohm's approach we avoid the notion of two different physical worlds and hence avoid all the paradoxes of Bohr's theory which ultimately are based on this notion. Also, we avoid the problem of quantum-to-classical transition as now classical physics appears to be merely a limiting case of quantum theory, which is relevant only insofar as physical objects may be regarded as being separable. This means that in contrast to the standard quantum theory, Bohm's theory is in a perfect harmony with the classical physics. However, this is not the case if one takes its relationship with the relativity and ultimately that is the reason why Bohm's theory is currently in the status of an apocryphal version of quantum theory. The point is that, to be recognized as the deepest model of universe, the quantum (Bohm's) model of universe should rest on a purely quantum particle dynamics beyond relativity. But no such dynamics is currently known even on the level of a conjecture. As a result, the quantum model may be regarded only as an addition to the relativistic model. This is precisely what was assumed by Bohm himself. However, even in the status of an addition, his model looks almost incompatible with relativistic kinematics.

The main difficulty is that if the characteristic lengthscale of quantum indivisibility has become macroscopic, then we inevitably come to what is known as quantum nonlocality. Today we know only one type of quantum indivisibility of a macroscopic lengthscale. It is the indivisibility of entangled distant particles. And the corresponding nonlocality was first noted by Einstein, Podolsky and Rosen (EPR) as long ago as in 1935 [9]. In those times, the very possibility of nonlocal correlations was regarded as an argument against the Bohr's version of quantum theory. But today it has become clear that nonlocality is actually an unavoidable attribute of any version. Therefore, the possibility of EPR nonlocality appears rather an argument in favor of Bohr's version at least in the context of its competition with the realistic version. Indeed, it is precisely the explicit mysticism of Bohr's theory which allows one to reconcile EPR nonlocality with the Minkowski model. By contrast, in the frames of Bohm's approach, EPR nonlocality implies the existence of *real* correlations between the past and the future and for most physicists this seems inacceptable even though EPR correlations are fundamentally non-causal. Perhaps this is the reason why Einstein took no enthusiasm for Bohm's theory despite the fact that, as such, this theory is a brilliant confirmation of his guess concerning the realizability of a realistic quantum theory with the so-called "hidden variable".

At present, the incompatibility of a realistic view of EPR nonlocality with the Minkowski model of spacetime is an agreed position of the majority of physical community. As a result, the so-called quantum dilemma is recognized which declares "either nonlocality or realism" [10]. Accordingly, when experimental evidence was provided for the EPR nonlocality, it was regarded as a strong argument against quantum realism. This evidence is based on the so-called Bell's theorem of 1964 where it was proved mathematically that a statistical processing of EPR measurements may answer the question whether or not nonlocal EPR correlations really exist

[11]. To date such measurements have been performed by several independent groups and they all show that nonlocal EPR correlations do exist precisely in accordance with quantum predictions [12-14].

However, there is a characteristic feature which makes EPR measurements to be hardly compatible with the Bohr's idea of two physical worlds. The point is that the lengthscale of EPR correlations may be as high as more than 100km [15]. At this lengthscale, it seems untenable to interpret EPR measurements in terms of a "non-existent" quantum world. In fact, there are only two alternative ways to resolve this problem. The first one is most consistently presented by Anton Zeilinger [16-17]. His interpretation of EPR nonlocality rests on the standard Bohr's theory but avoids the Bohr's idea of two physical worlds. In fact, Zeilinger puts forward the idea that the foundations of the entire physics are inherently mystical. This idea, however, seems incompatible with the very notion "science" and therefore may cause even a public protest (see, e.g., [18]).

At the same time, only one way is currently discussed which may be viewed as an alternative to the Zeilinger's approach. The way is that we should come back to the Lorentz-Poincare's version of relativity with its concept of a preferred frame of reference, the so-called "aether". In fact, soon after the realization of Bell test, this way was proposed by Bell himself as well as by Karl Popper and eventually was supported by Basil Hiley and David Bohm [19-21, 8]. Today, this way is becoming increasingly popular among physicists, including those who are directly engaged in EPR research [22]. On the one hand, along with the preserving of quantum realism, this way implies a new relativistic kinematics different from the Minkowski model and therefore it allows one to eliminate some well-known paradoxes of this model together with those which were found very recently (see, e.g., [23]). But, on the other hand, this way has a clear drawback. The point is that currently there is a number of mutually exclusive guesses about what could be called "aether." But none of them has been verified experimentally. Moreover, there are no ideas of how we could verify them even in principle. This means that to avoid the paradoxes of quantum mysticism as well as the paradoxes of Minkowski model, we come to such an almost mystical (and hence paradoxical) notion as unobservable "aether". This, as a minimum, looks inconsistently. Therefore, by default, the first ("no-aether") way is regarded as the most believable for the majority of physicists though only a few of them dare to bring this way to its logical conclusion, as was done by Zeilinger.

Finally, there may be a chance to make a choice between the Bohm's theory and the standard quantum mechanics without any appeal to the models of universe. This chance may arise if we will found a situation in which the predictions of the former theory differ from that of the latter. Certain efforts in this direction are currently undertaken but still without a success

[24]. The main problem is that such situations are very exotic and hence they are extremely difficult to implement as well as to interpret. Moreover, even if we obtain a result in favor of Bohm's theory, this would not clarify the current situation but rather would make it even more confusing because the problem of compatibility of Bohm's theory with relativistic kinematics remains unresolved.

## II.  Main part

### 2.1. An unexpected way to reconcile quantum realism with relativity

Nevertheless there may be a "third way" of how to avoid the paradoxes caused by the quantum mysticism as well as by the relativistic kinematics though it has never been discussed so far even at the level of a conjecture. This is because no one ever imagined that there could be a spatial dynamics beyond relativity. Therefore, despite many problems associated with the relativistic kinematics, it nevertheless remains in the status of the deepest kinematics ever known. Accordingly, the century-old taboo on a nonlocal signaling still remains. However, if a deeper-than-relativistic (quantum) dynamics nevertheless exists, then, instead of the status of an addition to Minkowski model, the Bohm's model would have the status of the deepest model of universe. In this case, it has the power to present its own concept of space and time without the taboo on a nonlocal signaling. And in searching for the purely quantum dynamics, it is the time to recall a prototype of the dynamics we need. We mean the electron transitions between atomic orbitals if we believe them to be realistic. Thus, the key question is whether or not spatially-discontinuous transitions are possible in macrocosm and, if yes, whether or not they can be observable.

Of course, it seems extremely unlikely that we will ever find a macro-system which is an exact copy of a single atom. But we can approach the problem in a more subtle way and try to find a macro-system with orbit-like quantum states of the lengthscale of system itself.

### 2.2. Spatially-separated macroscopic quantum states in the integer quantum Hall system

At first glance, our approach seems unpromising as it is well-known that if electron wavefunctions are macroscopic then, as a rule, they are not spatially-separated. This is the case in conventional solid-state systems with free electrons as well as in any superconducting materials. In both these cases, we are dealing with a system of indistinguishable electrons obeying quantum statistics. By contrast, in a single atom, electrons are spatially-separated and hence distinguishable. As a result, their behavior should be regarded on a non-statistical level.

Nevertheless, today a macro-system is known where electron wavefunctions are spatially-separated and may be of the lengthscale of the system itself. Such system is realizable in two-dimensional (2D) semiconductor quantum structures where the width of conducting layer is shorter than the electron mean free path. In these structures, the electron motion across the layer is quantized and therefore, instead of a continuous energy spectrum, there is a series of quantum levels. On the basis of these structures, one can realize the so-called integer quantum Hall (IQH) system or as it is sometimes called the quantum Hall state of matter [25-26]. In fact, it is precisely the system where a purely quantum effect manifests itself at a macroscopic lengthscale. We mean the integer quantum Hall effect, the discovery of which by Klaus von Klitzing was noted the Nobel Prize in 1985 [27].

To clarify the origin of spatially-separated macroscopic states in the IQH system, consider what happens if one applies a strong magnetic field perpendicular to 2D layer (*Z* axis). In this case, the electron motion is quantized not only in the *Z* direction but in the *XY* plane as well. As a result, each level related to the size quantization gives rise to a series of sublevels known as Landau levels. The energy of these levels is: $E_N = \hbar\omega_c (N+1/2)$, where N is the Landau quantum number, $\omega_c = eB/m^*c$ is the so-called cyclotron frequency (*m\** - the electron effective mass). In such system, most electrons are in the flat microscopic orbits with the radius $r = (\hbar c/eB)^{1/2}$. Nominally, the coordinate of the center of such orbit ($X_0$) is in a correlation with its wave vector along the *Y* axis: $X_0 = r^2 k_y$. But actually this correlation has no physical meaning because of the axial symmetry of the system in the *XY* plane. Thus, we have a strongly degenerated system ($E_N$ is independent of $X_0$) where the number of electron states per each Landau level is determined by the dense packing of electron orbits over the system area at a given magnetic field.

As we see, there are no quantum macro-states yet. But they immediately arise if one takes into account the electron behavior in a narrow strip (of the order of *r*) near the system edges [28]. In this strip, there is a strong in-plane electric field perpendicular to the edge. As a result, electrons are in a crossed electric and magnetic field. In terms of classical physics, they should thus drift along the equipotential lines near the edges. However, to describe their behavior more adequately, one needs to apply quantum formalism. This was first done by Robert Laughlin, and then, in more details, by Bertrand Halperin [29-30]. According to their calculations, instead of the classical trajectories, now we have orbit-like quantum states with a characteristic width of the order of *r*. This means that along with a number of micro-orbits, there are a few spatially-separated macro-orbits near the system edges (**Fig. 1a**). Such macro-orbits are known as *extended states* and these are precisely the states responsible for the integer quantum Hall effect which manifests itself through an exact quantization of transverse conductivity (along the *Y* axis) when an electric bias is applied along the *X* axis.

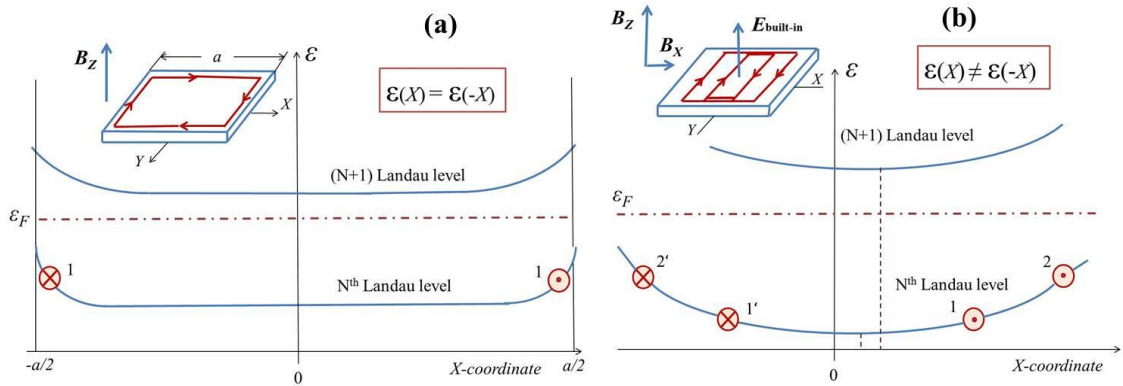

**Fig. 1. Macroscopic orbit-like quantum states in IQH-type systems**

**(a) Energy spectrum of conventional (symmetric) IQH system.** Landau level degeneracy is lifted near the sample edges (*a* – the sample length) due to the presence of an electric field pointed toward the sample center. In a combination with the external perpendicular magnetic field, it gives rise to spatially-separated orbit-like quantum states. One of these states is shown schematically in the figure. In this state, electron behaves as a spontaneous current flowing around the macroscopic sample. Dash-dot line denotes the Fermi energy. Inset shows schematically the spatial configuration of macroscopic orbits.

**(b) Estimated energy spectrum of an infinite IQH system with asymmetric confining potential when external magnetic field has both quantizing and in-plane components.** Landau level degeneracy is now lifted throughout the whole system due to a combined effect of both "built-in" electric field and in-plane magnetic field. As a result, electrons are in spatially-separated quantum states in which they behave as spontaneous currents flowing in opposite directions along the *Y* axis. We show two pairs of such currents and in each pair the currents are a counterpart of each other. Spatial asymmetry manifests itself through a microscopic spatial shift of Landau levels, which increases with increasing of the Landau quantum number. Vertical dashed lines show bands' minima. Inset shows schematically the expected spatial configuration of macro-orbits in a real finite system.

Thus, the structure of an ordinary IQH system is truly reminiscent of that of a single atom in the sense that this system also has orbit-like quantum states of the lengthscale of system size. Nevertheless, it is clear that even if we will provide an electron transition between these macro-orbits, the electron will not overcome a macroscopic distance as the orbits are shifted from each other by a microscopic distance.

## 2.3. Spatially-discontinuous electron transitions through an intermediate orbit-like macroscopic state: A gedanken experiment

However, the very fact of macroscopic orbit-like states may open the door to a peculiar type of spatially-discontinuous transitions which is unrealizable in the microcosm. We mean the transitions where macroscopic orbit-like state plays the role of an intermediate state for the spatially-discontinuous transition between two distant local levels. In this case, spatial discontinuity results from the fundamental indivisibility of the intermediate state.

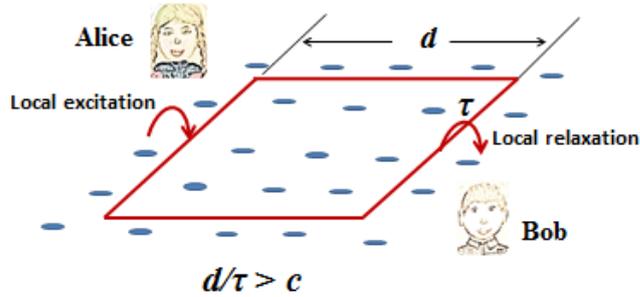

**Fig. 2.** *Gedanken* **experiment to show a potentiality for a spatially-discontinuous electron transition between two distant local states through an intermediate orbit-like macroscopic state**
Here a macroscopic spatial discontinuity of the transition is due to the fundamental indivisibility of the intermediate macroscopic orbit-like state. *d* – the length of the side of a macroscopic square orbit, *τ* – electron lifetime at the orbit. Solid cycles show schematically charged local scatterers.

To clarify this idea, consider a *gedanken* experiment under the conditions very close to the real IQH system (**Fig.2**). Suppose there is an orbit-like quantum state of a square shape with macroscopic side *d*. The left-hand side of the square will be called the region *A* (Alice's) while the opposite side – the region *B* (Bob's). As in any real IQH system, the orbit is surrounded by point-like charged scatterers which transfer electron from the orbit into a nearby local state during a characteristic time *τ*. Suppose we inject electron from a low-energy local state into the orbit by the light quanta that excite the region *A*. Eventually, the electron will thus appear in a local state near the orbit due to the scattering. But the crucial point is that the probability of being localized in the region *B* is *exactly the same* as the probability of being localized in the region *A*. It is precisely due to the fundamental indivisibility of the orbit or, in terms of Bohr's theory, due to the fundamental "non-existence" of electron in the orbit before the scattering which plays the role of a measurement.

Thus, during the time *τ*, the electron will overcome the distance *d* with the probability of one fourth. But it is easy to see that *τ* is generally *independent* of *d* as it is merely a characteristic of the material. Therefore, we potentially can enlarge the orbit and hence increase the distance *d* to fulfill the following inequality: $d/\tau > c$, where *c* is the speed of light. Moreover, if one takes a typical (for IQH systems) scattering time (about 3ps) as well as a typical sample length (about 1cm), the ratio $d/\tau$ will be one order *higher* than the speed of light. Also, we can deliberately increase the concentration of scatterers in Bob's region (say, through a doping) providing thus a much lower $\tau_p$ as well as a much higher probability for the electron localization in this region. Therefore, if our *gedanken* experiment is realizable in an IQH-type system, then the inherent nonlocality of spatially-discontinuous dynamics could manifest itself in a real experiment.

In fact, our *gedanken* experiment demonstrates a profound physical meaning which lies in the fact that quantum formalism appeals not to the real space but rather to a multi-dimensional Hilbert space. Indeed, the indivisibility of a pure quantum state (such as a macroscopic orbit) directly follows from the fact that this state is characterized by an indivisible point (or ray) in Hilbert space whereas this point may correspond to a macroscopic region in the real space. Historically, Einstein was the first who regarded the appeal to Hilbert space as a serious problem. He noted this problem on the general debate at the Solvay Conference of 1927. Perhaps even at that distant time he suspected that this point may put at stake his principle of locality. So far, however, his intuitive suspicions have not been confirmed and only now, when an atom-like quantum object appears macroscopic, we see that they were not groundless.

*2.4. Asymmetric IQH system as a system with atom-like pervasive quantum ordering*

**2.4.1. The structure of macroscopic orbit-like states in an asymmetric IQH system: A guess**

At first sight, we can perform our *gedanken* experiment in the ordinary IQH system. But actually there is a serious difficulty. As we noted, here the number of macro-orbits is relatively small and they all are concentrated in a very narrow strip near the edges. As a result, to detect an effect related to these orbits, we need a method which would be sensitive to them but insensitive to a much higher number of microscopic states. This requirement is fulfilled in von Klitzing's magneto-transport experiments because macroscopic currents cannot be provided by microscopic states. However, no other such methods are known yet. Perhaps that is the reason why the IQH effect is named after Edwin Hall who first measured transverse conductivity back in the nineteenth century. Thus, if we want to implement our *gedanken* experiment, then we should first find the method of how to deal with the macro-orbits.

To solve this problem, it should be noted that actually the emergence of macro-orbits is related to the fact that electrons appear in a gigantic crossed electric and magnetic field which is inaccessible in the bulk. However the very structure of 2D systems prompts us the method of how we could access such conditions not only near the edges but in the whole IQH system. Indeed, 2D systems are precisely the systems where electrons may undergo a gigantic electric field without being accelerated because of the quantization of their motion along the $Z$ axis. Moreover, several technological methods are known of how to provide a gigantic electric field across the 2D layer. One of them is that we could make the surface-to-well distance as low as a few tens of nanometers in order the so-called surface potential (of about 0.8eV) could easily penetrate into the well giving rise to a gigantic electric field of the order of $10^5$V/cm [31]. In the literature, such field is often called "built-in" field ($E_{\text{built-in}}$) and the structures where this field is high enough are often called the asymmetric 2D structures. And this field gives us a chance to provide a high enough crossed electric and magnetic field in the whole IQH system. To this aim, we should only provide the external magnetic field which has not only a quantizing component ($B_z$) but also an in-plane component ($B_x$).

Generally speaking, the calculations of energy spectrum of an infinite 2D system in presence of crossed electric and magnetic field have already been done by Gorbatsevich *et al.* [32-33]. They have shown that the Landau level degeneracy may truly be lifted in such a way that Landau levels transform into something like energy bands where electrons behave as spatially-separated currents flowing in opposite directions along the *Y* axis. Their characteristic width is of the order of the cyclotron radius related to the in-plane magnetic field ($r_x$) and their position along the *X* axis is again determined by the familiar relation $X_0 = r_x^2 k_y$. However, the point is that here this relation has a well-defined physical meaning as the *X* axis is now precisely the direction of the in-plane magnetic field. Thus, the system is such that each one-electron current has a spatially-separated "counterpart" which flows in strictly opposite direction and is characterized by the same energy as well as by the same absolute value. An asymmetry of the energy spectrum manifests itself through a small mutual shift of the Landau bands. Schematically, such spectrum is shown in **Fig. 1b**. Now, if we shift to a real finite system, then it seems reasonable to assume that each current together with its counterpart is actually the same macro-orbit extended along the *Y* axis and closed near the sample edges parallel to the *X* axis. In this case, each Landau band should consist of a great number of almost rectangular macro-orbits where the longer high-energy orbits are closer to the edges parallel to the *Y* axis. Schematically, the spatial distribution of macro-orbits is shown in the inset to **Fig. 1b**.

The above speculations clearly cannot guarantee that macro-orbits could truly be realized in the whole asymmetric IQH system. At best, they show *a potentiality* for this. Moreover, there is no guarantee that the macro-orbits (if any) are detectable in a real experiment. Therefore, prior to an attempt to realize our *gedanken* experiment, we should carry out preliminary experiments to demonstrate unambiguously the presence of such states and their accessibility. In fact, these experiments should answer the question whether or not there may be a quantum phase transition which is induced by crossed electric and magnetic field and results in the transformation of an ordinary IQH system into a peculiar quantum phase with spatially-ordered macro-orbits distributed over the whole system.

To design preliminary experiments, it should first be noted that if a 2D system is truly filled with macro-orbits, then our experimental method may differ from the von Klitzing's method. More specifically, instead of the external electric bias, now we could excite the system by light quanta precisely like in our *gedanken* experiment. However, as for the detection of system's responses, it seems expedient to measure again the macroscopic currents (here along the *Y*-axis) precisely because they cannot be provided by any microscopic states.

Phenomenologically, such currents may be induced by light due to the system's asymmetry related to a nonzero vector product $B_x \times E_{\text{built-in}}$. Nominally, it would be the so-called photo-voltaic effect which, as a rule, is detectable under an intense terahertz laser excitation of asymmetric semiconductor systems [34]. However, if we are truly dealing with a system of spatially-separated macro-orbits, then the photo-voltaic effect should differ in principle from that observed in the system of indistinguishable electrons. Indeed, if the energy spectrum is such as in **Fig. 1b**, then the system's translational symmetry is broken along the *X* axis because spontaneous currents depend strongly on the *X* coordinate in each Landau band and moreover these bands are shifted from each other. This means that even if the photo-excitation is strictly homogeneous, light-induced local currents along the *Y* axis may nevertheless be a strong function of *X* coordinate and moreover their dependence on the magnetic field may not be the same in different local regions. Most likely, such effect (if any) should manifest itself under the so-called cyclotron resonance (CR) conditions when the energy of light quanta is such that the electron could transit between the neighboring Landau bands with different electron occupancy. And if a strong difference between the local responses does occur, then it would be a strong argument that we are truly dealing with macro-orbits distributed over the whole system.

### 2.4.2. Experimental evidence for the atom-like spontaneous quantum ordering

As a starting material to prepare asymmetric IQH system, we use the so-called GaSb-InAs-GaSb single quantum well structures grown by the method of molecular beam epitaxy (MBE). The width of the InAs conducting layer is 15nm and it is sandwiched between two thin AlSb barriers (3nm each) to avoid the so-called hybrid electron-hole states. Such structures are known to have a high electron sheet density (about $2 \cdot 10^{12} \text{cm}^{-2}$) as well as a high concentration of charged point-like defects responsible for quasi-elastic scattering. For this reason, the scattering time is as short as about 3ps and the electron mean free path is as low as about 0.1μm. To provide an asymmetry of confining potential, the surface-to-well distance is as low as 20nm so that the surface potential can easily penetrate into the well giving rise to the "built-in" field of about 100kV/cm [35].

As we see, the selected system is such that it has no macroscopic parameters with the dimension of length. This means that if we observe a macroscopic quantum effect, then its characteristic lengthscale may only be of the order of system size in the *XY* plane. In our experiments, this size is 20mm along the *X* axis and 16mm along the *Y* axis and this size cannot yet be significantly increased for technological reasons. To detect light-induced local currents along the *Y* axis, we use four pairs of short ohmic contacts, which are shifted from each other along the *X* axis (**Fig. 3a**). The length of each contact is 2mm, the distance between the neighboring pairs is 3mm, and the distance between the contacts in the same pair is 12mm. To avoid a direct effect of sample edges, all contacts are shifted from the edges by at least 1.5mm.

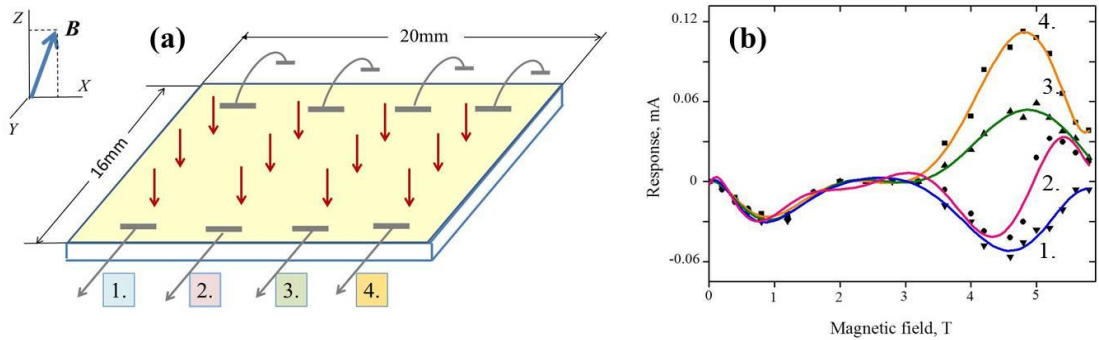

**Fig. 3. Evidence for a** spontaneous **quantum ordering throughout the whole system**

**(a)** Sketch of the experiment. All sample area is exposed to the spatially-homogeneous laser radiation. Light-induced electric responses are detected through four contact pairs shifted from each other along the $X$-axis

**(b)** The outcome of the experiment. Solid lines are a guide for the eyes. Curves are numbered in accordance with the numbering of the contact pairs.

Experiments are performed at low temperature (about 1.9K) when the thermal energy is much less than the typical energy of Landau quantization. To provide a nonzero in-plane component, external magnetic field is tilted from the normal by about $15°$. The field is provided by a superconducting magnet (up to 6T). As a source of light quanta, we use pulsed ammonia laser optically pumped by $CO_2$ laser. The laser wavelength is 90.6μm so that the energy of light quanta is about 13.7meV. In the selected system, CR conditions are thus expected in quantizing magnetic field of about 4.6T. Under these conditions, the interband electron relaxation is provided by the so-called optical phonons and its characteristic time is about 30ps, i.e. one order longer than the characteristic time of intraband relaxation due to the quasi-elastic scattering [36]. To avoid considerable heating, the laser operates in a single-pulse regime with the pulse duration as short as about 50ns. Light intensity is of the order of 200W/cm$^2$. High-speed electric responses are detected in a short-circuit regime with a 50Ω load resistance per each channel of the detection. Time resolution is about 10ns.

**Fig. 3b** shows dependence of light-induced responses on the magnetic field for each contact pair. It is clearly seen that there are two domains where the responses exist. In the low-field domain (up to 2T), the Landau quantization can be neglected. Therefore, we are dealing with a system of indistinguishable electrons and, as expected, here the responses are almost identical. The high-field domain lies in a wide range around the CR point and here the responses do differ drastically from each other so that even their resonant curves are different as well and may be either monopolar or bipolar. Such a qualitative difference between the responses clearly cannot be provided by long-range Coulomb potential, especially if one takes into account that the screening length is microscopic in any real IQH systems.

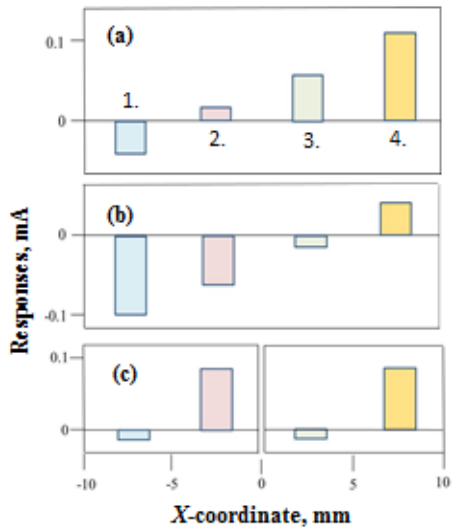

**Fig. 4. Evidence for an indivisible behavior of the system under study**

(a) Local responses at $B = 5T$; (b) Local responses after reversing the magnetic field; (c) Local responses after splitting the sample

Thus, our first experiment supports the assumption regarding spatially-separated macro-orbits in the whole system. But, to avoid any ambiguity, we reverse magnetic field and repeat the measurements in both domains. In the low-field domain ($B = 0.8T$), responses behave quite trivial again: each one merely changes its sign. But in the high-field domain ($B = 5T$), their behavior differs even in principle. To show this, we present the spatial diagrams of the responses for two opposite directions of magnetic field (**Figs. 4a** and **4b**). As we see, the reversal of magnetic field leads to a rotation of the diagram *as a whole* by $180°$. This means that our system does behave as an indivisible whole so that such fundamental symmetry relation as the oddness with respect to magnetic field is fulfilled only on the level of the whole system but is not necessarily fulfilled on the level of an individual response.

To demonstrate an indivisible behavior of our system more directly, we mechanically split the sample along the *Y* axis into two identical parts and measure local responses in each fragment. At $B = 0.8T$, responses are insensitive to the splitting, as expected. However, this is not the case at $B = 5T$ (**Fig. 4c**). Comparing with **Fig. 4a**, we see that *all* responses have changed drastically *regardless* of their remoteness from the splitting line. Moreover, all these changes are such that the spatial diagrams of responses become the same in both parts and each one is qualitatively reminiscent of the diagram of the initial system. Finally, we reverse the magnetic field and, as before, this leads to a rotation of each new diagram by 180° about the center of the corresponding fragment.

Thus, all experiments confirm the presence of a quantum ordering of the lengthscale of system size. To some extent, our system may thus be regarded as a toy model of undivided universe with a pervasive quantum ordering. However, there is a principle difference between the ordering we have faced and the pervasive ordering implied by Bohm. In our case, the ordering is related not to the quantum entanglement but rather to a spatial structure of the wavefunctions of *individual* quantum particles like in the case of a single atom. And in macrocosm, this type of quantum ordering may lead to a deeper nonlocality than the nonlocality provided by the entanglement-related quantum ordering. Indeed, in the former case, a strictly local perturbation of an ordered system may be detectable far away from this perturbation and moreover this detection may be done on a non-statistical level. In fact, this characteristic feature of the single-atom-like quantum ordering has already manifested itself in the experiment with a mechanical splitting where we see a purely quantum transfer of strictly local perturbation at a macroscopic distance due to a re-ordering of macroscopic electron wavefunctions throughout the whole system. And this transfer is clearly detectable on a non-statistical level. However, although, in itself, this effect seems non-trivial, it nevertheless is unrelated to the spatially-discontinuous dynamics which actually is the only sufficient argument to abandon the relativistic kinematics and to come to a deeper quantum kinematics. Therefore, to achieve our ultimate goal, we proceed to the realization of our *gedanken* experiment by using of our macroscopic system with an atom-like pervasive quantum ordering.

## 2.5. Observation of a spatially-discontinuous electron dynamics

In fact, our previous experiments give us all the necessary tools to implement our *gedanken* experiment. First of all, they have shown that light-induced electron transitions between macro-orbits may lead to such local responses which may serve as an indicator of non-equilibrium electrons in a given region. Also, when developing the scheme of the main test, we assume a microscopic spatial shift between the Landau bands so that the macro-orbits are supposed to be almost symmetrical with respect to the sample center. With this in mind, we take a standard sample with only two contact pairs symmetrical with respect to the sample center. These are precisely the pair No.1 and the pair No.4 in the previous experiments. The first region (pair No.1) we will call the region *A* while the second region (pair No.4) we will call the region *B*. Thus, we will repeat the experiment shown in **Fig. 3a** but with a principle difference, that is, now we cover the sample with a non-transparent plate which has only two windows, in the region *A* and in the region *B*, so that each of these regions may be either illuminated or not (see the left-hand panel of **Fig. 5**).

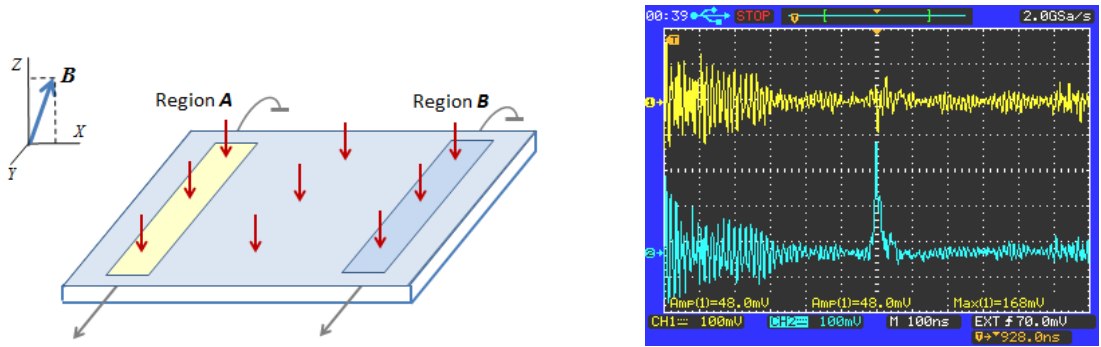

**Fig. 5. Observation of light-induced electron dynamics with a macroscopic spatial discontinuity**

**(Left-hand panel)** Sketch of the test. Conditions are the same as in **Fig. 3a** but now there are only two contact pairs and the sample is covered by a non-transparent plate with two windows: in the region *A* (formerly pair No. 1) and in the region *B* (formerly pair No. 4). Responses are measured in both regions. In the figure, only the window *A* is open.

**(Right-hand panel)** Typical tracks when only the window *A* is open ($B = 5T$): upper track – response from the *illuminated* region *A*; lower track – response from the *distant* (from the laser spot) region *B*. Both responses are equally pre-amplified. Timescale is 100ns/div.

It is easy to see that our test is truly an implementation of the gedanken experiment. The only difference is that now we have a great number of spatially-separated macro-orbits as well as a great number of exciting light quanta. Also, electrons will most likely be excited not from local levels but rather from macro-orbits belonged to the lower Landau band. Likewise, the electron scattering will most likely be accompanied by the transition not to a local level but rather to the other macro-orbit which, however, now belongs to the same Landau band. Nevertheless, all these differences are rather of a quantitative character and may affect only the sensitivity of our test.

As such, our test is extremely transparent. At the first stage, we measure light-induced responses in both regions (*A* and *B*) when both windows are open. At the second stage, we perform the same measurements when only one window is open (either *A* or *B*). As before, we start with indistinguishable electrons ($B = 0.8T$) and, as before, the outcome is quite trivial. At the first stage, when both windows are open, we see the same responses in both regions while, at the second stage, when only one window is open, response occurs only in the illuminated region. After that we shift to the spatially-separated electrons ($B = 5T$). At the first stage, when both windows are open, the outcome seems quite trivial as well: responses occur in both regions and the ratio between them is precisely the ratio between the response No.1 and the response No.4 under the full illumination (see **Fig. 4**).

However, what we observe at the second stage seems amazing or even incredible at least in terms of our everyday (or "classical") intuition. Indeed, when we close the window *B*, the response *B* does *not* disappear. Instead, it decreases only by half despite the fact that the region *B* is now at a macroscopic distance from the lase spot. Moreover, the response A also decreases by half. This means it is sensitive to an event that happened at a macroscopic distance from its region A. In other words, after the closing of the window B, the ratio between the responses remains the same as if both regions are still equally illuminated but now with a halved intensity. It follows that the number of excited electrons is the same in both regions despite the fact that one of them is at a giant distance from laser spot, which is five orders longer than the electron mean free path. Furthermore, if we close the window A and open the window B, both responses remain the same. This means that the illumination of only one region (either A or B) provides the same number of excited electrons in both regions. But this is absolutely impossible if electrons reach the distant region through a continuous motion from the illuminated region because a gigantic distance (with respect to their mean free path) should thus be overcome without any loses. Moreover, it is easy to see that such a continuous transport of electrons without loses remains impossible even if there is no any scattering at all because there are no reasons for which all electrons move precisely toward the distant region but not in any other direction.

Now, to estimate experimentally the temporal characteristics of the electron transport, we carry out a synchronous detection of both responses when only the window A is open. Typical tracks are presented in the right-hand panel of Fig. 5. It is seen that the ratio between the responses is indeed the ratio detected under the full illumination (see Fig. 4). This means that under a certain orientation of magnetic field, the distant response may be even three times higher than the response in the illuminated region. But even more importantly, we see no a delay between the responses at least with an accuracy of about 10ns. At the same time, it is easy to estimate that even if there is a delay of less than 10ns, then, to overcome the distance from A to B in a proper time, electrons should move with a speed as high as more than $10^8$cm/s. It is really too fast transport even if it is of a ballistic character without any scattering.

Finally, to demonstrate explicitly the lengthscale of the spatial discontinuity, we perform a one more experiment. We close the window B and then shift the window A toward the window B in order to measure both responses when we illuminate only an intermediate region between A and B. It turns out that, as soon as the window leaves the region A, both responses completely disappear and then appears once again only when the window reaches the region B. This fact clearly indicates that the spatial discontinuity is truly of the lengthscale of system size and therefore it has no a fundamental upper limit.

## III. Fundamental Consequences

Thus, our experimental test shows unambiguously that spatially-discontinuous electron transitions can truly be realized in macrocosm and they have no fundamental limitations on the maximum lengthscale of discontinuity. Therefore, they can be regarded as a peculiar quantum dynamics beyond relativity. Actually, the very fact of such dynamics clearly indicates that relativity cannot exhaustively describe any particle dynamics in macrocosm as this theory is based on the postulate of continuity of any spatial dynamics which always can meaningfully be characterized by the notion "speed". It follows that the relativistic model of universe can no longer be regarded as the deepest model. Such model should be provided by the theory which is capable of describing this new dynamics. This is precisely the quantum theory which now must inevitably be realistic since only realistic theory can provide a macroscopic model of universe. Therefore, we come to the de Broglie-Bohm quantum theory which leads to the Bohm's model with its non-mechanistic ("quantum") understanding of the structure of universe, which differs drastically from the current "relativistic" understanding.

As it was noted above, the core of the quantum understanding is that the so-called elementary particles can no longer be regarded as fundamental entities which are a kind of "building blocks" of the universe. Quite the contrary, the only fundamental entity is precisely the entire universe so that, in general, any consideration of its constituent parts separately from an overall system is always a kind of approximation. With this approach, any elementary particle appears to be rather an emerging object. Therefore, the current trend of an increasingly deeper fragmentation of matter in searching for a deeper "basis" of the universe is inherently flawed as we always will be dealt with emerging physical objects those ultimately are not fundamental entities. As Bohm repeatedly emphasized, quantum laws require the abandon of the current understanding of the universe as being a complex mechanism with a huge number of "sub-mechanisms" which are self-sufficient and in turn can always be decomposed into "sub-sub-mechanisms" and so on. Instead, if we regard quantum laws realistically, then we come to a fundamentally new insight of the universe which implies that "*… each element that we can abstract in thought shows basic properties ... that depend on its overall environment, in a way that is much more reminiscent of how the organs constituting living beings are related, than it is of how parts of a machine interact…*" [8].

As is known, until now the major problem of Bohm's model of universe is that it has no its own particle dynamics beyond relativity. Therefore, it could not claim to have the status of the deepest model of universe. At best, it could only claim to have the status of an addition to the relativistic model. This is precisely the status which was supposed by Bohm himself. However, even in this status, his model looks incompatible with the Minkowski model, especially after the realization of Bell test. But now, when a purely quantum particle dynamics has been demonstrated, Bohm's model has all the grounds to be recognized as the deepest model of universe. Accordingly, now we can remove the centenary-old taboo on a nonlocal signaling because now such signaling has nothing to do with causality. Moreover, the possibility of a nonlocal signaling directly follows from the very fact of the spatially-discontinuous dynamics as it is inherently nonlocal. Also, through this dynamic we ultimately come back to the concept of absolute simultaneity. However, this is a quantum concept of absolute simultaneity, which differs in principle from the classical concept as the former goes beyond the postulate of continuity of any spatial dynamics and therefore beyond the notion "speed".

As it is easy to see, the observation of quantum dynamics disavows a fundamental character of the notion "speed" as this notion appears to be not an inevitable link between the truly fundamental categories such as space and time. Accordingly, it disavows a fundamental character of the Lorentz invariance which ultimately is based on the notion "speed". As a result, the term "nonlocality" loses its fundamental meaning as well. Thus, instead of the relativistic concept of spacetime, we come back to the well-known classical understanding of space and time. However, such a revival of both classical space and classical time is now based on a deeper quantum insight of universe where the notion "speed" is no longer fundamental.

In fact, as it now becomes clear from the quantum model of universe, the main inconsistence of relativistic kinematics is that it actually postulates a fundamental (self-sufficient) character of what we call "physical objects". Only in this case a spatially-discontinuous dynamics of these objects is fundamentally impossible. In fact, this subtle point was noted by Einstein himself so that in his autobiographical notes we can read the following: "*The theory [special relativity] ... introduces two kinds of physical things, i.e., (1) measuring rods and clocks, (2) all other things, e.g., the electro-magnetic field, the material point, etc. This, in a certain sense, is inconsistent; strictly speaking measuring rods and clocks would have to be represented as solutions of the basic equations (objects consisting of moving atomic configurations), not, as it were, as theoretically self-sufficient entities*" [37]. But, as we see, quantum theory denies the full self-sufficiency of individual physical objects and this is especially important for microscopic objects whose wavefunctions are actually a characteristic of their "non-self-sufficiency". Therefore, the lengthscale of their functions is precisely the lengthscale, within which we can face such a "mystical" thing as their spatially-discontinuous dynamics. And insofar as this lengthscale may be macroscopic, their spatially-discontinuous dynamics may also be macroscopic. Actually, this is precisely what we have demonstrated experimentally.

In fact, the highest status of quantum model of universe automatically implies a new hierarchy of fundamental physical theories, where the de Broglie-Bohm realistic quantum theory appears the deepest physical theory. And this hierarchy ultimately allows us to achieve such a long-awaited result as the removal of any contradictions between the quantum theory and the relativity precisely because the latter no longer defines our view of kinematics. As for the dynamics, it is easy to see that here the predictions of quantum and relativistic theories do not actually overlap as they are relevant to different conceptual levels of reality. This is precisely the reason why it seems meaningless to talk about a unification of quantum and relativity theories. Rather, we should talk about a unification of relativity and classical physics as conceptually these theories are at the same level and their predictions can truly be comparable. From this comparison we see that the relativistic dynamics is actually a refinement of classical dynamics and this refinement is especially important when the speed of a continuous dynamics approaches the speed of light. Therefore, these theories may well be regarded as the same (unified) theory which is relevant only insofar as physical objects may be regarded as fundamental, self-sufficient entities.

On the other hand, if we remain within the framework of relativistic kinematics and therefore within the framework of Bohr's quantum mechanics, then quantum theory will still be inconsistent with relativity even at the level of their philosophical foundations as the former is essentially a mystical theory while the latter is strictly realistic. It seems this is precisely the reason why almost century-long efforts to unify these theories have not actually lead to a success. Today this problem is known as the problem of so-called "unified theory" or, as it is sometimes called, the "theory of everything" and this problem is still regarded as one of the most vital problems facing the modern physics (see, e.g., [38]). And only now it becomes clear why this problem is unsolvable in the frames of traditional approaches.

However, the problem of "unified theory" is not the only problem that can be resolved in the framework of the new insight of physical world. The other problem which is solved automatically is the problem of interpretation of the quantum theory itself because, as it was noted above, the de Broglie-Bohm's version of quantum theory is actually the only version provided a complete quantum model of universe, which is consistent with the spatially-discontinuous particle dynamics. This fact, however, does not detract from the merits of Bohr's mathematical formalism which still remains a robust method to predict the outcome of various experiments, at least those of them which are realizable to date. The best confirmation of the relevance of Bohr's formalism lies in the fact that, as such, macroscopic spatially-discontinuous transitions are fully consistent with it. In fact, Bohr's formalism may be regarded as a limiting case of a deeper (Bohm's) quantum formalism and this limiting case is such that it does not imply any interpretation. By default, this view of Bohr's formalism is currently shared by most physicists who (deliberately or not) follow the well-known principle "shut-up-and-calculate" [39]. And actually this principle does express in the best possible way of how we should view of

Bohr's formalism although there was an attempt to construct a model of universe precisely on the basis of a pure mathematics without what might be called an interpretation [40].

A one more aspect of the quantum view of universe is that it helps us to solve not only physical problems but also some epistemological problems arose in the context of Bohr's quantum mechanics. Historically, these problems were the subject of a lengthy discussion between Bohr and Einstein, which today is known as Bohr-Einstein debate. As is known, throughout all his life Einstein insisted that "*… without the belief that it is possible to grasp reality with our theoretical constructions, without the belief in the inner harmony of our world, there could be no science…*" [41]. On the contrary, Bohr believed that quantum theory compromises the very notion of explanation of physical phenomena as well as the possibility of a philosophical comprehension of any physical theory. Instead, he used the notion of description of physical phenomena and ultimately insisted that quantum theory should be isolated from any philosophy. Today the majority of physicists agree that the clear success of the predictions of Bohr's quantum mechanics automatically implies that we should make our choice in favor of the Bohr's view.

But as we come to quantum realism, we thus rehabilitate the Einstein's view that any physical theory must have a realistic interpretation and therefore can be philosophically comprehended. Also, we rehabilitate the principle which was regarded by him as the foundation of all science. It is the principle of knowability of outside world, the adherence to which is still regarded as his deep misconception. Moreover, through the quantum view of universe, this principle surprisingly acquires a much deeper meaning than Einstein originally intended. This is related to the fact that, in a sense, the very notion of absolute simultaneity excludes such purely relativistic term as the "hopelessly distant objects". This means that the potentiality for exploring the outside world is much higher than we thought so far. Finally, it should be noted that potentially there may be much larger objects with the structure reminiscent of that of a single atom (see, e.g., [42]). And this may open up new horizons to study macroscopic systems with a pervasive quantum ordering.

To conclude, one would recall an intuitive guess expressed by Bell as long ago as back in 1986 when discussing the problems related to the compatibility of quantum and relativity theories in the light of the demonstration of the EPR nonlocality. This guess is as follows: "I think it is very probable that the solution to our problems will come through the back door; some person who is not addressing himself to these difficulties with which I am concerned will probably see the light…" [19]. Now it seems we have found this "back door" and the light is really seen.

# Acknowledgements

The MBE samples were kindly provided by Prof. Sergey Ivanov (Ioffe Institute).